\documentclass[final,3p,times,twocolumn,fixfloat]{elsarticle}
\usepackage{graphicx}% eps fig, package for more complicated commands
\usepackage{amssymb} % various useful mathematical symbols
\usepackage{amsmath} % extended theorem environments

\usepackage{lineno}
\usepackage{float} % improved interface for floating objects
\usepackage{subfig} % supersedes the subfigure package
\usepackage{ifpdf}
\ifpdf
\usepackage[pdftex]{hyperref}
\else
\usepackage[hypertex]{hyperref}
\fi

\hypersetup{
  pdftitle={},%
  pdfauthor={},%
  pdfsubject={},%
  pdfkeywords={},%
  pdfstartview={},%
  bookmarksopen=true, breaklinks=true, debug=true, %psi-gamma-290508.pdf
  colorlinks=true, linkcolor=red, citecolor=blue, urlcolor=blue
}

\journal{Physics Letters B: SLAC-PUB-13528}

\usepackage{amsmath}
\usepackage{epsfig}
\usepackage{float}
\usepackage{verbatim} 
\usepackage{color}

\newcommand{\ie}{{i.e. }}
\newcommand{\eg}{{e.g.}}

\newcommand{\Q}{{\cal Q}}

\newcommand{\cf}[1]{{Fig.~\ref{#1}}}

\begin{document}

%%%%%%%%%%%%%%%%%%%%%%%%%%%%%%%%%%%%%%%%%%%%
%% FRONTMATTER
%%%%%%%%%%%%%%%%%%%%%%%%%%%%%%%%%%%%%%%%%%%%

\begin{frontmatter}

%\leftline{} \rightline{SLAC-PUB-....}
\title{Real next-to-next-to-leading-order QCD corrections to $J/\psi$ and $\Upsilon$ hadroproduction in association with a photon}

\author[SLAC]{J.P.~Lansberg}\fnref{email}
\address[SLAC]{SLAC National Accelerator Laboratory, Theoretical Physics, Stanford University, Menlo Park, CA
94025, USA.}
\fntext[email]{Email address: {\tt lansberg@slac.stanford.edu}}

%\date{}

\begin{abstract}
\small
We update the study of the QCD corrections to direct $J/\psi$ and $\Upsilon$
hadroproduction in association with a photon in the QCD-based
approach of the Colour-Singlet (CS) Model. After comparison with
the recent  full next-to-leading-order (NLO) computation for this process, 
we provide an independent confirmation to the inclusive case
that NLO QCD corrections to quarkonium-production processes whose LO exhibits a non-leading $P_T$ behaviour 
can be reliably computed at mid and large $P_T$ by considering only the real emission contributions
accompanied with a kinematical cut. In turn,  we evaluate the leading part of 
the $\alpha^4_S\alpha$ contributions, namely those coming
from $(J/\psi,\Upsilon)+\gamma$ associated with two light partons.  We find that they
are dominant at mid and large $P_T$. This confirms our expectations from the leading
$P_T$ scaling of the new topologies appearing at NNLO. We obtain that the yield from the 
CS becomes one order of magnitude larger than 
the upper value of the potential Colour-Octet yield. The polarisation of the $^3S_1$ quarkonia produced in 
association with a photon is confirmed to be longitudinal
at mid and large $P_T$ . 
\end{abstract}


\end{frontmatter}

\section{Introduction}

For a long time, the many difficulties to correctly predict quarkonium-production 
rates at hadron colliders have been attributed to 
non-perturbative effects associated with channels in which the heavy
quark and antiquark are produced in a colour-octet state~\cite{Lansberg:2006dh,Brambilla:2004wf,Kramer:2001hh}.
On the basis of leading-order (LO) ($\alpha_S^3$) calculations, it had been assumed
that colour-singlet production channels give a small contribution at mid
and large $P_T$. The confusion most probably came 
from the fact that quantum-number conservation ($J$, $P$, $C$,
and colour) prevents leading $P_T$ scaling at LO and, in glue-glue
production, at next-to-leading order (NLO) ($\alpha_S^4$)\footnote{except for the 
subprocess $gg\to \Q + Q \bar Q$.}. Thus, in contrast with the
situation for many other observables, there is still a possibility
of unexpectedly large (colour-singlet) contributions at large $P_T$ at
next-to-next-to-leading order (NNLO) ($\alpha_S^5$).
In addition, two unexpected features of fragmentation approximations 
 have recently been revealed: heavy-quark fragmentation only dominates 
 over the other topologies at very high $P_T$ and gluon fragmentation may not
dominate over  double $t$-channel gluon exchanges. As a consequence, it seems that, 
when leading kinematic contributions are
correctly accounted for, colour-singlet production channels will play a
more important role --if not the most-- than in past analyses  in the explanation of
quarkonium-production observables and  this suggests in turn that colour-octet fragmentation
contributions may be less important than had previously been thought.

\begin{figure*}[ht!]
\centering
\subfloat[]{\includegraphics[scale=.35]{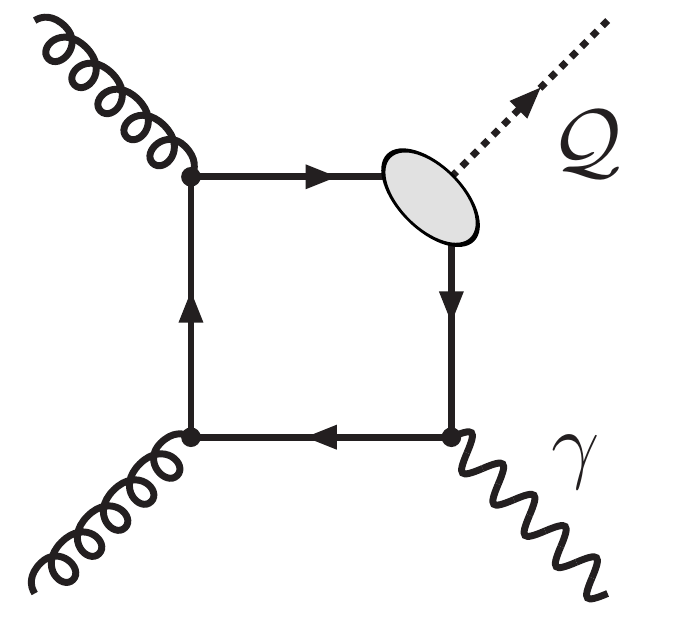}}
\subfloat[]{\includegraphics[scale=.35]{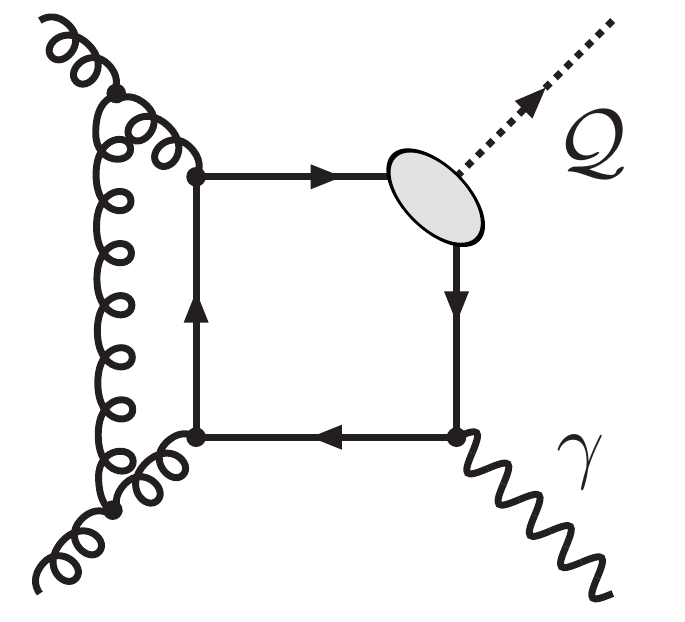}}
\subfloat[]{\includegraphics[scale=.35]{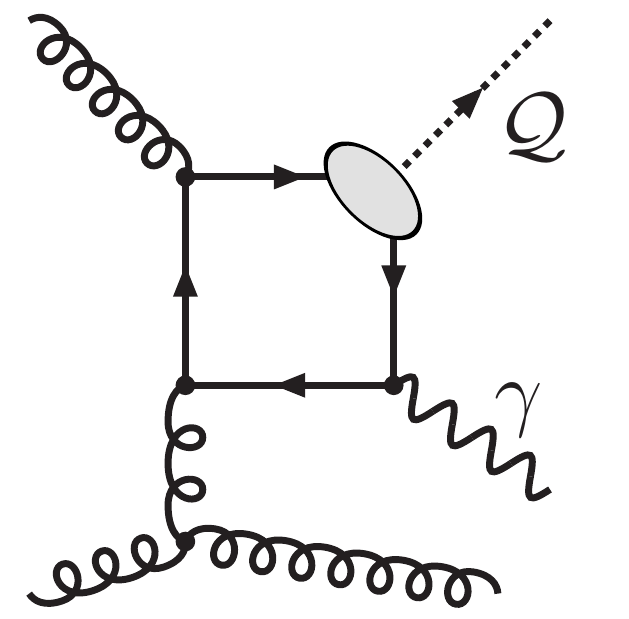}}
\subfloat[]{\includegraphics[scale=.35]{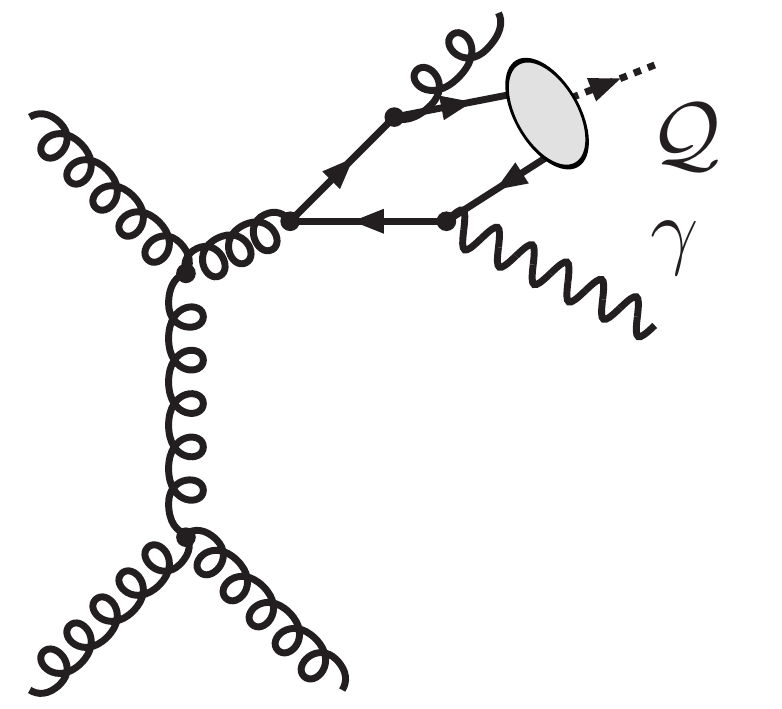}}
\subfloat[]{\includegraphics[scale=.35]{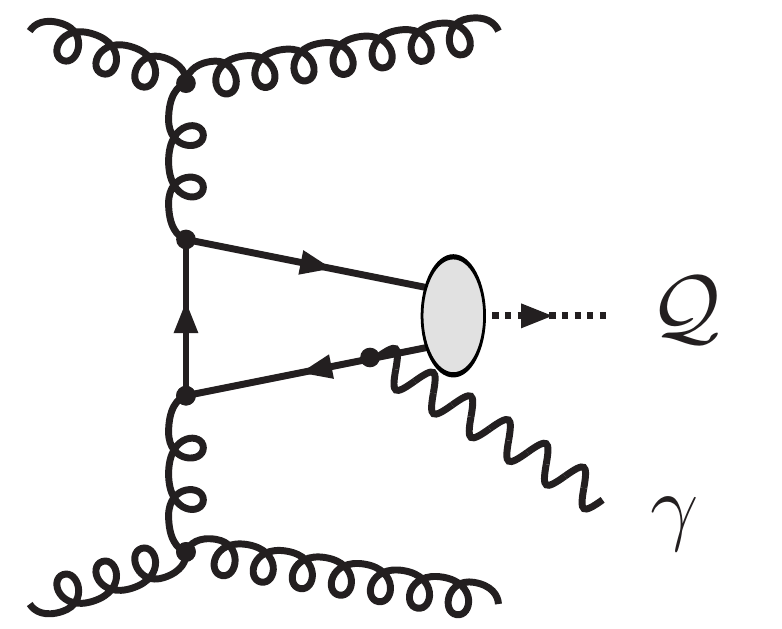}}
\subfloat[]{\includegraphics[scale=.35]{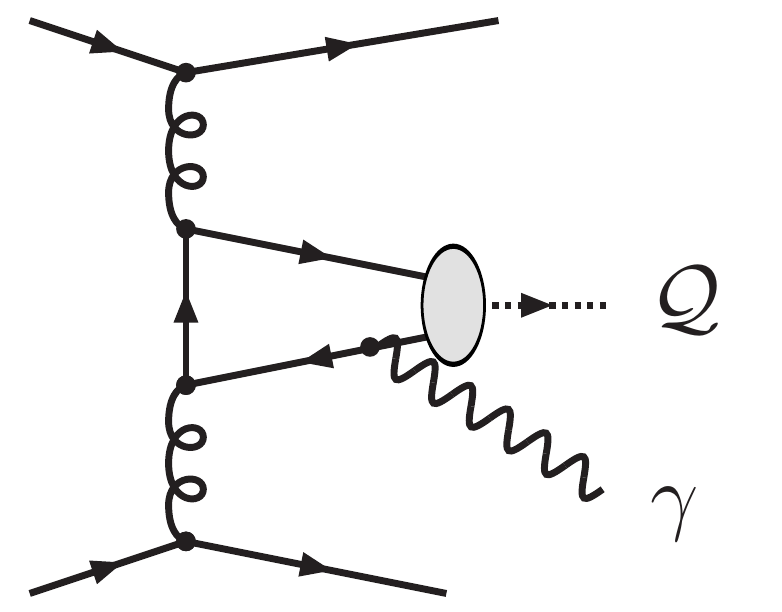}}
\caption{Representative diagrams contributing to the hadroproduction of a $J/\psi$ in association
with a photon at orders $\alpha_S^2 \alpha$ (a), $\alpha_S^3 \alpha$ (b,c), $\alpha_S^4  \alpha$ (d,e,f). See discussions in the text.}
\label{diagrams}
\end{figure*}

This became clear thanks to the several recent computations 
of QCD corrections to quarkonium hadroproduction processes.
The NLO  corrections to the inclusive yield
of $J/\psi$ and  $\Upsilon$ were computed \cite{Campbell:2007ws,Artoisenet:2007xi}
in the QCD-based approach of the Colour-Singlet (CS) Model~\footnote{
The CSM can be also regarded as the leading order contributions in the heavy-quark-velocity ($v$) expansion
of the effective theory, Non-Relativistic Quantum Chromodynamics (NRQCD)~\cite{Bodwin:1994jh}.}~\cite{CSM_hadron}.
Its polarisation was in turn computed \cite{Gong:2008sn} at NLO. These computations were recently
complemented \cite{Artoisenet:2008fc} by the addition of the real
NNLO corrections --thereafter referred to as NNLO$^\star$--. It was then shown that there may be no need 
to incorporate Colour-Octet (CO) transition (higher-$v$ corrections of NRQCD)
to describe the hadroproduction of $\Upsilon$ at the Tevatron~\cite{Affolder:1999wm,Acosta:2001gv,Abazov:2005yc,Abazov:2008za}. 
In the case of the $J/\psi$ and $\psi'$~\cite{pierreHLPW08,Lansberg:2008gk}, the CS contributions are significantly enhanced 
and brought very near the experimental data of CDF although the large $P_T$ direct yield seems not to be fully accounted for in 
the case of the $\psi'$ for instance. As regards the CO channels for $J/\psi$ production, their NLO QCD corrections 
were recently computed in \cite{Gong:2008ft}. It was seen that they minimally affect both the $P_T$ dependence 
and the normalisation of the partonic matrix elements at mid- and large-$P_T$, thus also the value of the $^3S_1^{[8]}$ CO 
Long-Distance 
Matrix Elements (LDME) fit to the data. 
Similarly, the polarisation prediction are not modified and remains in disagreement with the latest CDF
measurements~\cite{Abulencia:2007us}.  For recent reviews, the reader is guided 
to~\cite{Lansberg:2006dh,Brambilla:2004wf,Kramer:2001hh}
along with some perspectives for the LHC \cite{Lansberg:2008zm} and 
a recent discussion on those aforementioned QCD correction computations~\cite{Lansberg:2008gk}.

Beside the studies of inclusive production, efforts are being made to obtain improved theoretical predictions
for complementary observables to the inclusive yield, such as hadroproduction of $J/\psi$
 and $\Upsilon$ (thereafter commonly named $\Q$) in association with a 
photon~\cite{Berger:1993sj,Doncheski:1993dj,Kim:1994bm,Mirkes:1994jr,Roy:1994vb,Kim:1996bb,Mathews:1999ye,Kniehl:2002wd}. Recently,
the NLO corrections to the hadroproduction of $\Q+\gamma$
  in the CS channel  were computed by Li and Wang in \cite{Li:2008ym}. 
As in the inclusive case, NLO corrections are significant in the large $P_T$ region since new topologies
appear with slower $P_T$ falloff (see \cf{diagrams} (c)), in comparison to LO topologies \cf{diagrams} (a)
and other NLO topologies as the loop corrections \cf{diagrams} (b).

   In \cite{Artoisenet:2008fc},  it was shown
that  some $\alpha^5_S$ contributions to the inclusive yield coming from three-jet configurations, 
\ie ${\cal Q} + jjj$, such as those arising from gluon-fragmentation and
``high-energy enhanced'' (or double $t$-channel gluon exchange)  
channels are important and clearly dominate over the other contributions at mid and large $P_T$.
In the case which interests us here, $\Q+\gamma$, those two channels appear at order $\alpha^4_S\alpha$ -- see for instance 
\cf{diagrams} (d) and  (e,f)-- and are expected to dominate over the $\Q+\gamma$ yield at mid and large $P_T$.

In this work, we therefore apply the same procedure as in \cite{Artoisenet:2008fc} to evaluate their contributions. First, we will
check that the NLO computation of Li and Wang \cite{Li:2008ym} can indeed be reproduced by an evaluation of the 
 real $\alpha_S^3\alpha$ contributions  ($\Q+\gamma$ + one light partons) complemented by a cut-off on low 
invariant masses for any pairs of external light partons in the process. The procedure is explained in the next section.
We will also check that the sensitivity of our computation on this cut-off dies away when $P_T$ grows and that the theoretical
uncertainty attached to its choice is typically smaller than the ones attached to the choices of the renormalisation scale $\mu_r$
for $\alpha_S$, the factorisation scale $\mu_f$ for the (collinear) parton distribution functions (PDF) 
and the heavy-quark mass $m_Q$.

Having performed those checks, we will apply the same procedure for the evaluation of the contribution for 
$\Q+\gamma$ + two light partons --namely the real NNLO contributions to $\Q+\gamma$ production-- arguing that they provide with a first reliable estimate
of the complete NNLO contributions to $\Q+\gamma$ production at large enough $P_T$. As regards the polarisation, 
the quarkonia directly produced in association with a photon via those channels
are mainly longitudinally polarised, as in the inclusive case.

\section{Cross section at NLO}

\subsection{Inclusive case}

As we argued in \cite{Artoisenet:2008fc}, the NLO contributions to the inclusive yield
 can be approximated at large enough $P_T$ in a relatively simple and reliable manner by computing the 
$\alpha^4_S$ contributions consisting in the production of a $\cal Q$ with 2
light partons (denoted $j$ thereafter) on which we apply a cut-off on low 
invariant masses for any light parton pairs in the process. Computations of such cross sections
 can be done reliably  using  the automated generator of matrix elements MadOnia~\cite{Artoisenet:2007qm}.

\begin{figure*}[ht!]\centering
\subfloat[$J/\psi + X$]{\includegraphics[width=.99\columnwidth]{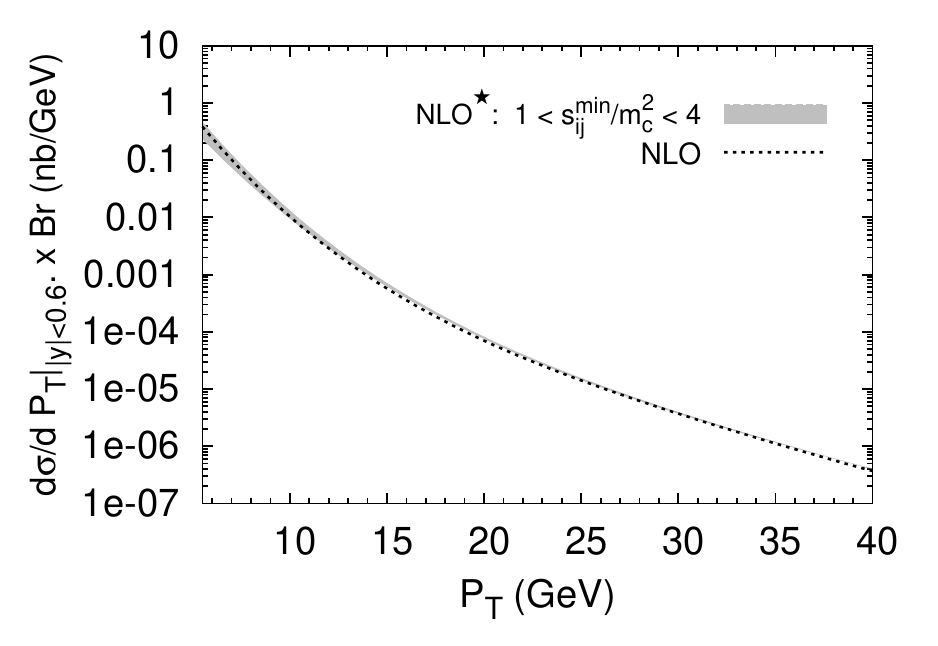}}
\subfloat[$\Upsilon + X$]{\includegraphics[width=.99\columnwidth]{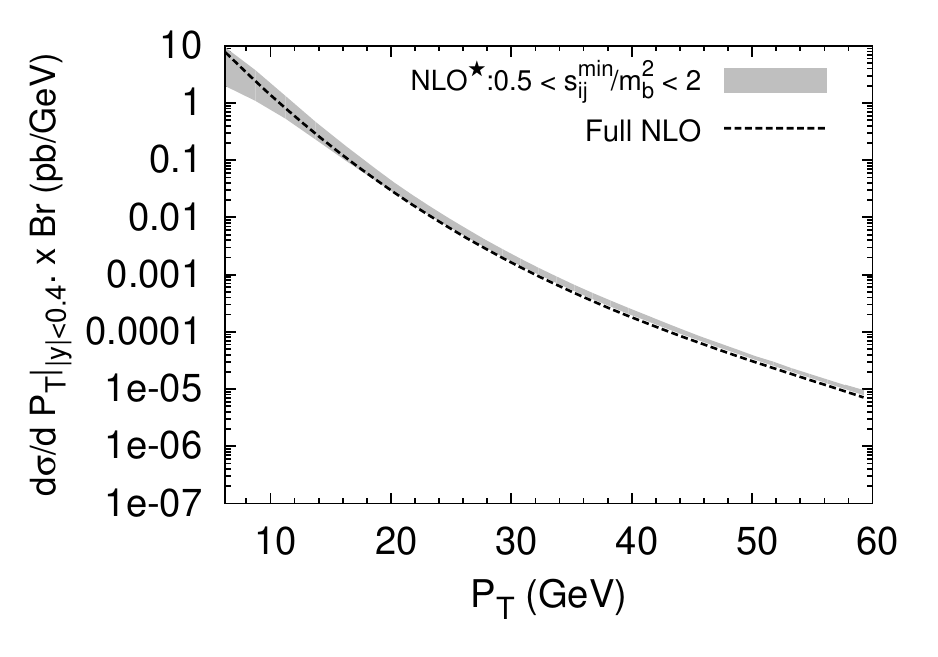}}
\caption{(a) full computation at NLO for  $J/\psi + X$ (dashed line)~\cite{Campbell:2007ws}  
vs. NLO$^\star$ ($J/\psi$ + 2 light partons with a  cut on $s_{ij}$) (gray band) at $\sqrt{s}=1.96$ TeV; (b) full 
computation at NLO for  $\Upsilon(1S) + X$ (dashed line)~\cite{Campbell:2007ws}  vs. NLO$^\star$ 
($\Upsilon(1S)$ + 2 light partons with a cut on $s_{ij}$) (gray band) at $\sqrt{s}=1.96$ TeV. See text  for details, {\sf Br} stands for
the respective branching into dileptons.
} \label{fig:NLO_vs_NLO_star_inclusive}
\end{figure*}

\begin{figure*}[ht!]\centering
\subfloat[$J/\psi + \gamma+ X$]{\includegraphics[width=.99\columnwidth]{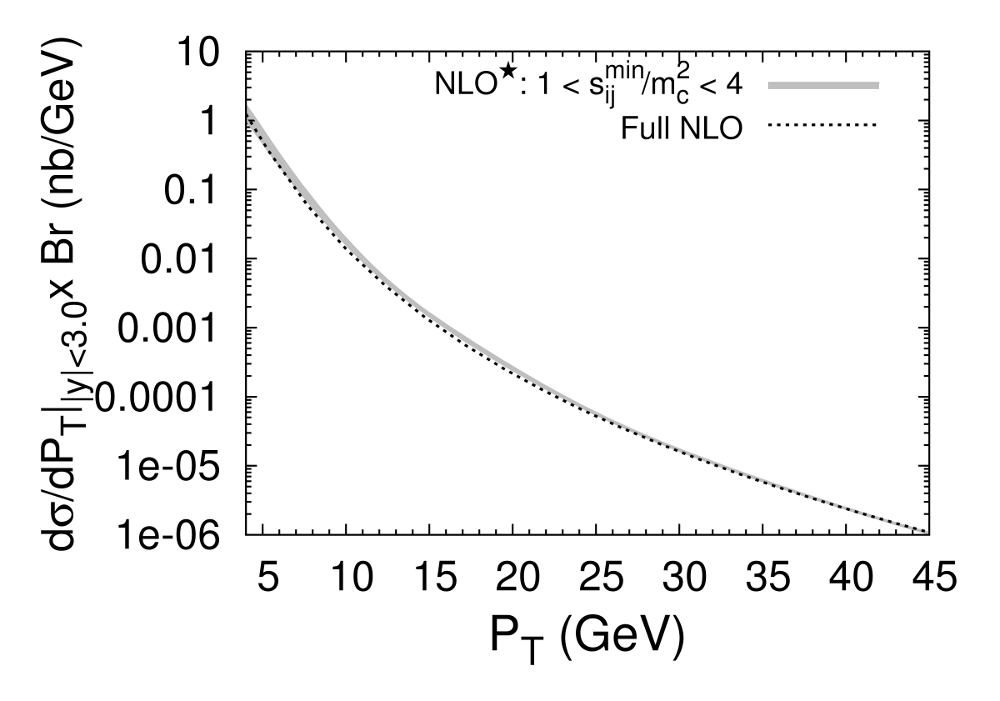}}
\subfloat[$\Upsilon + \gamma+ X$]{\includegraphics[width=.99\columnwidth]{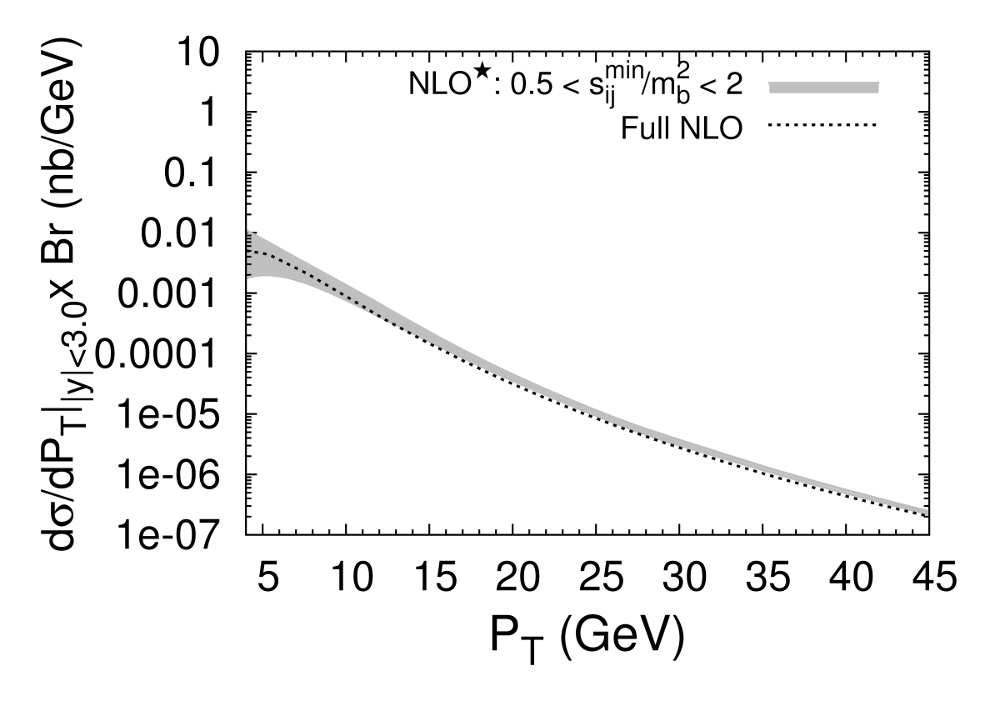}}
\caption{(a) full computation at NLO for  $J/\psi +\gamma+X$ (dashed line)~\cite{Li:2008ym} 
 vs. $J/\psi +X$ + 1 light parton with a  cut on $s_{ij}$ (gray band) at $\sqrt{s}=14$ TeV; (b) 
full computation at NLO for  $\Upsilon(1S) +\gamma+ X$ (dashed line)~\cite{Li:2008ym}  
vs. $\Upsilon(1S) +\gamma$ + 1 light parton with a
 cut on $s_{ij}$ (gray band) at $\sqrt{s}=14$ TeV. The absolute value of 
the rapidity of both the $\Q$ and the $\gamma$ is limited to 3.
} \label{fig:NLO_vs_NLO_star}
\end{figure*}

The underlying idea supporting this was twofold:
\begin{itemize}
\item  First, at large enough $P_T$, topologies with
  the leading $P_T$ behaviour will dominate and those are wholly included in this subset of $\alpha^4_S$ contributions 
(the production of a $\cal Q$ with 2 light partons); 
\item   Second, this subset accounts for a physical process at Born level.  
Its contribution is therefore finite except  for soft and collinear divergences. The purpose of the cut-off is to 
avoid such divergences by imposing a lower bound on the invariant-mass of 
any light parton pairs ($s_{ij}$). For the new channels (with a leading $P_T$ scaling) opening up at $\alpha^4_S$,
 the dependence on this cut gets 
smaller for large $P_T$ since no collinear or soft divergences can appear there.
For other channels, whose Born contribution is at $\alpha^3_S$,
the cut would produce logarithms of $s_{ij}/s_{ij}^{\rm min}$. Those can be 
large. Nevertheless, they can be factorised over their corresponding Born
contribution, which scales at most as $P_T^{-8}$. The sensitivity on 
$s_{ij}^{\rm min}$ is thus expected to vanish at large $P_T$.
\end{itemize}

This insensitivity to the cut and the good agreement between the NLO$^\star$ (${\cal Q}+jj$ with a $s_{ij}$ cut) and the full NLO 
result is recalled in \cf{fig:NLO_vs_NLO_star_inclusive}  for the case of the inclusive $J/\psi$ and $\Upsilon(1S)$  production. 
The gray band illustrates the sensitivity to the invariant-mass cut 
$s_{ij}^{\rm min}$ between any pairs of light partons when it is varied from 
$m_c^2$ to $4m_c^2$ and $0.5 m_b^2$ to $2m_b^2$.  In both NLO and NLO$^\star$ computations, the value of all parameters
were set to the same values. For the $J/\psi$, we have  $m_c=1.5$ GeV, $|R(0)|^2=0.810$ GeV$^3$, 
$\mu_f=\mu_r=\mu_0=\sqrt{4m_c^2+P_T^2}$ and Br$(J/\psi \to \mu^+\mu^-)=0.0588$ and,  for the $\Upsilon(1S)$,
 $m_b=4.75$ GeV, $|R(0)|^2=6.48$ GeV$^3$, $\mu_f=\mu_r=\mu_0=\sqrt{4m_b^2+P_T^2}$ and Br$(\Upsilon\to \mu^+\mu^-)=0.0218$.
The parton distribution set used was CTEQ6\_M~\cite{Pumplin:2002vw}. The yield becomes 
insensitive to the value of 
$s_{ij}^{\rm min}$ as $P_T$ increases, and it reproduces very accurately the 
differential cross section at NLO accuracy, both for the $J/\psi$ and $\Upsilon(1S)$ case.

\subsection{Production in association with a photon}

In the more exclusive case ${\cal Q}+\gamma$, similar topologies are present 
with the same $P_T$ scaling and we also expect to reproduce accurately the 
yield at NLO accuracy ($\alpha^3_S\alpha$) computed in \cite{Li:2008ym} by 
computing the yield from the production of ${\cal Q}+\gamma$ with
one light parton with the invariant-mass cut $s_{ij}^{\rm min}$ between any pairs of light partons,
also referred to as NLO$^\star$. 

This is indeed the case. For instance,  the differential cross section 
for $\Q+\gamma$ at NLO accuracy from Li and Wang~\cite{Li:2008ym} 
is displayed in \cf{fig:NLO_vs_NLO_star}  and is very well reproduced by
the  NLO$^\star$ computed for different values of $s_{ij}^{\rm min}$. 
When $P_T$ grows, the latter becomes completely insensitive to the value 
chosen for $s_{ij}^{\rm min}$. The same parameter values were used for \cf{fig:NLO_vs_NLO_star}
as for \cf{fig:NLO_vs_NLO_star_inclusive} and $\alpha$ was set to $1/137$.

This result is a clear and completely independent confirmation of the validity of the reasoning
initially given in~\cite{Artoisenet:2008fc} that NLO QCD corrections to quarkonium-production 
processes whose LO shows a non-leading $P_T$ behaviour can be reliably computed at mid and large 
$P_T$ by considering only the real emission contributions
accompanied with a kinematical cut. In turn, this reinforces 
our confidence that the impact of NNLO contributions can evaluated likewise by computing the 
NNLO$^\star$ contributions, as done in the following section.

This also gives us confidence that  much better Monte Carlo simulations of inclusive production at
mid and large $P_T$ could be achieved using NLO$^\star$ and NNLO$^\star$ partonic matrix elements, which can be 
interfaced~\cite{Alwall:2008pm} with event-generators such as PYTHIA~\cite{Sjostrand:2006za}. In any case, 
they would give results much more reliable than simulations based
on matrix elements for CS channels at LO only.

\section{Cross section and Polarisation at NNLO$^\star$}

\subsection{Cross section}

Among the contributions appearing at $\alpha^4_S\alpha$, we find the topologies 
of~\cf{diagrams} (d) (gluon fragmentation) and~\cf{diagrams} (e,f) 
(``high-energy enhanced'' or double $t$-channel gluon exchange), those exhibit new kinematical enhancements 
appearing in higher-order QCD corrections. In other words, those  provide us with new
mechanisms to produce a high-$P_T$ ${\cal Q}$ with a $\gamma$ with a lower kinematic
suppression, still via CS transitions. They are therefore expected to dominate the
differential cross section at NNLO accuracy in the region of large
transverse momentum.

Those  are also entirely contained in the contributions to $pp \to {\cal Q} +\gamma+ jj$, namely
the real $\alpha_S^4 \alpha$ corrections, and we can follow the procedure validated in the previous section, 
by ``simply'' adding one light partons in the final state.

The computation of $pp \to\Q +\gamma+ jj$ at tree level 
is in principle systematic, but technically quite challenging: 
a dozen parton-level subprocesses contribute, most 
involving a few hundred  Feynman diagrams.  As done in~\cite{Artoisenet:2008fc}, 
we follow the approach described in Ref.~\cite{Artoisenet:2007qm}, which
allows the automatic generation of both the subprocesses and 
the corresponding scattering amplitudes. 

The differential cross-sections for $J/\psi +\gamma$ and $\Upsilon(1S) +\gamma$
are shown in~\cf{fig:dsdpt_NNLO_star}. The gray band  (referred to as
NLO$^\star$) corresponds to the sum of the LO and the real $\alpha_S^3 \alpha$ contributions. 
The red (or dark) band (referred to as NNLO$^\star$) corresponds to the sum of the LO, the real $\alpha_S^3 \alpha$ and 
the real $\alpha_S^4 \alpha$ contributions.  The  $\alpha_S^4 \alpha$  contributions 
in both case dominate over the yield at large $P_T$. The uncertainty bands are obtained from the combined variations
 $0.5 \mu_0 \leq \mu_{r,f}\leq 2 \mu_0$ with for the $J/\psi$, $m_c=1.5\pm0.1$ GeV  and 
$1 \leq s_{ij}^{\rm min}/(1.5~{\rm GeV})^2 \leq 2$ and,
for the $\Upsilon(1S)$,   $m_b=4.75\pm0.25$ GeV and $0.5 \leq s_{ij}^{\rm min}/(4.5~{\rm GeV})^2 \leq 2$.

\begin{figure}[ht!]\centering
\includegraphics[width=1.\columnwidth]{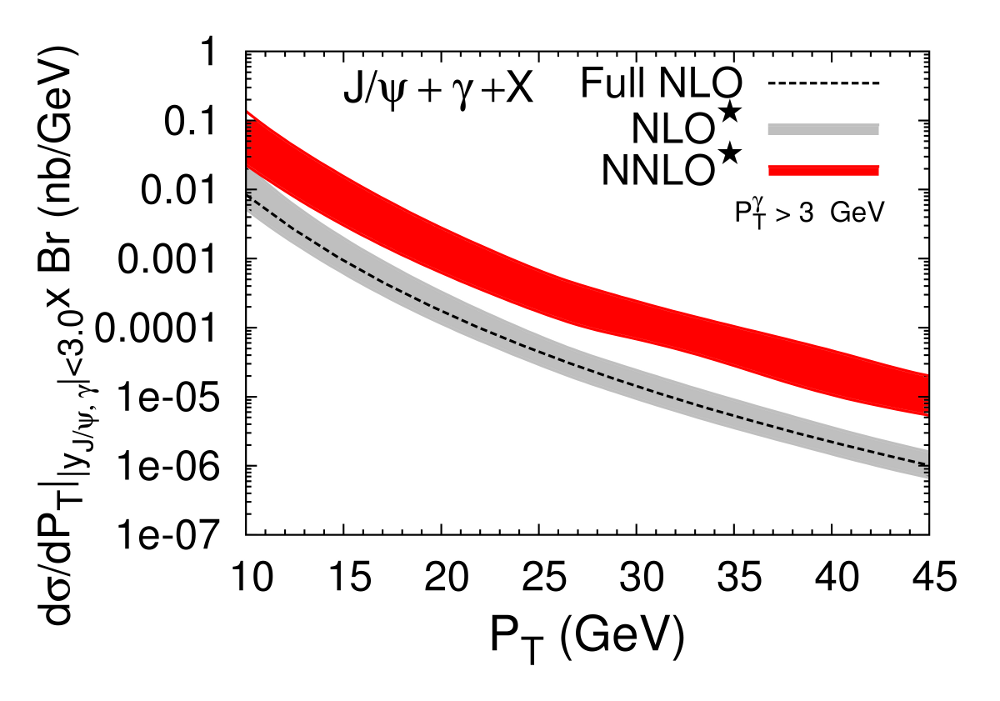} \\
\includegraphics[width=1.\columnwidth]{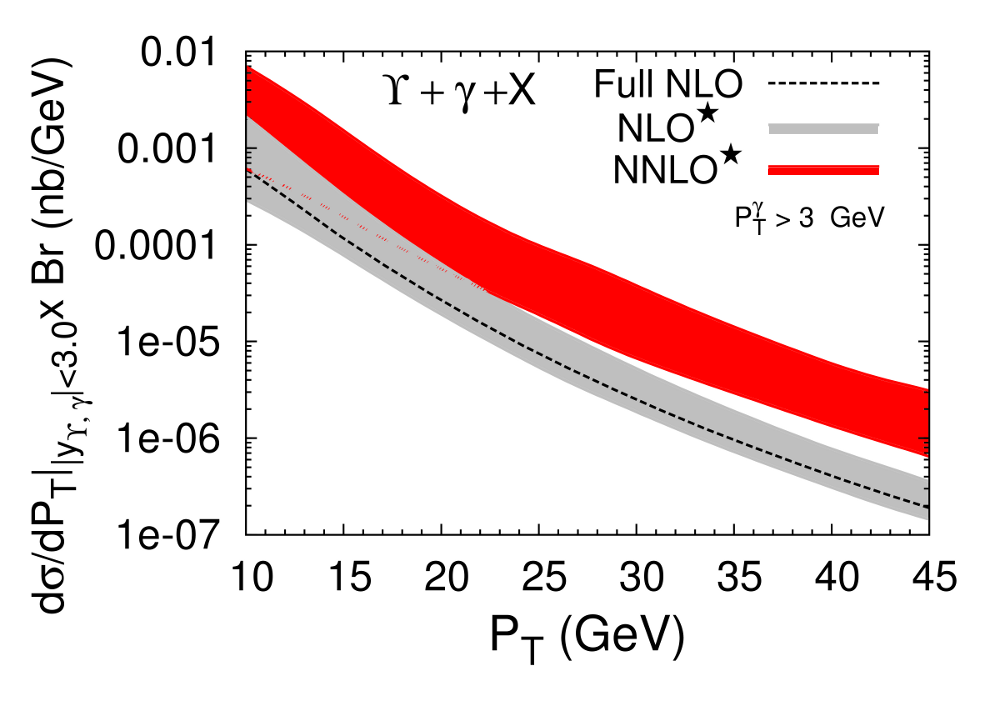}
\caption{
Results for (up) $J/\psi+\gamma$ and (down) $\Upsilon(1S)+\gamma$ from full NLO, 
NLO$^\star$ and NNLO$^\star$ contributions at $\sqrt{s}=14$ TeV.  
The theoretical-error bands 
for NLO$^\star$ and NNLO$^\star$  come from combining the
uncertainties resulting from the choice of $\mu_f$, $\mu_r$, $m_q$ and $s_{ij}^{\rm min}$. The absolute value of 
the rapidity of both the $\Q$ and the $\gamma$ is limited to 3, $P_T^\gamma > 3$ GeV.
A photon isolation cut ($\Delta R>0.1$) was applied on the NLO$^\star$ and NNLO$^\star$ yields (see discussion in 
section 5.)} \label{fig:dsdpt_NNLO_star}
\end{figure}

At the leading order in the heavy-quark velocity ($v$), the results for the radially excited states $\psi(2S)$, $\Upsilon(2S)$ and 
$\Upsilon(3S)$ are readily obtained by changing $|R_{\Q}(0)|^2$ and the branching
ratio into dileptons.

\subsection{Polarisation}

As regards the  polarisation parameter $\lambda$, it is computed by
analysing the angular distribution  ($\theta$) between the $\ell^+$ direction in the
quarkonium rest frame and the quarkonium direction in the laboratory
frame. The normalised angular distribution $I(\cos \theta)$ then reads
\begin{equation}
\label{angulardist}
I(\cos \theta) =
\frac{3}{2(\lambda+3)} (1+\lambda \, \cos^2 \theta)\,,
\end{equation}
from which we can extract $\lambda$  bin by bin in $P_T$.

\begin{figure}[ht!]
\centering
\includegraphics[width=0.95\columnwidth]{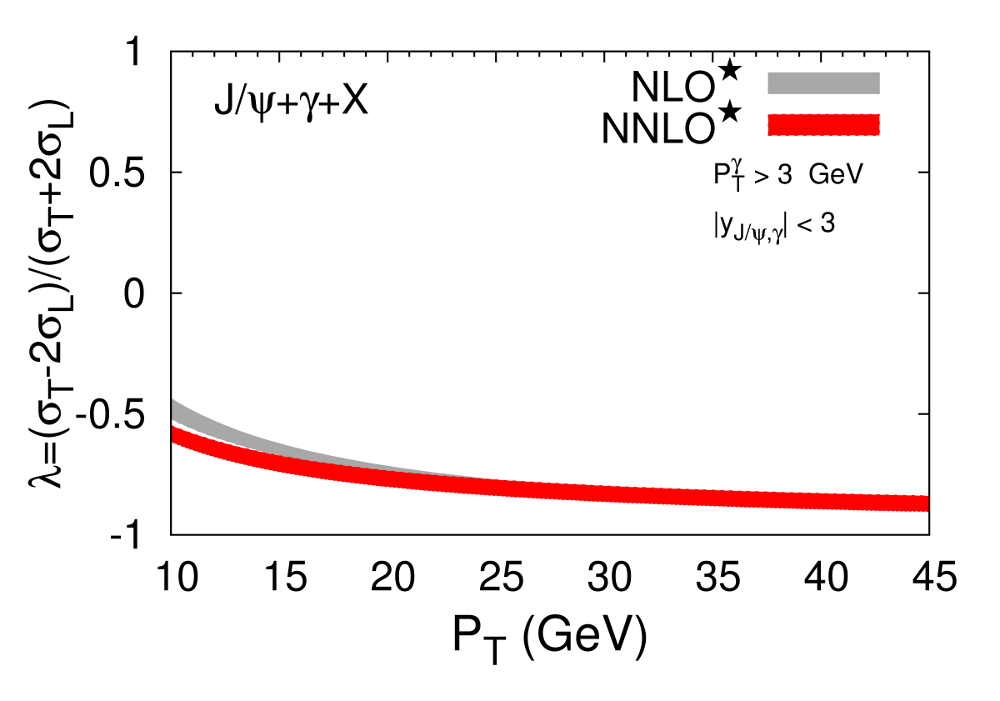}
\caption{Polarisation of the $J/\psi$ produced in association with a photon at $\sqrt{s}=14$ TeV 
up to the order $\alpha_S^4 \alpha$ (NNLO$^\star$). Most of the 
uncertainties on $\lambda$ from the choice of $m_c$ and $\mu_r$ cancel. The uncertainty band of the NLO$^\star$ and NNLO$^\star$ 
result comes from the variation of the cutoff $s_{ij}^{\rm min}$. }
\label{polarizationalphas4}
\end{figure}

Our results for the $J/\psi$ are shown in~\cf{polarizationalphas4}
along with the curves  for the NLO$^\star$. 
Our predictions for the polarisation parameter $\lambda$ for the NLO$^\star$
are in qualitative agreement with those of~\cite{Li:2008ym}, i.e. the
$J/\psi$'s produced in association with a photon are dominantly longitudinal. We did not go
further in the comparison since, for both numerical computations, 5\% precision in $\lambda$
could only be reached at a high cost of computing time and since the NNLO$^\star$
is anyhow larger. As regards the latter, it confirms the trends
of the NLO$^{\star}$ (and thus NLO) results. This is not a surprise knowing the NNLO$^\star$
results for the inclusive yield, differing essentially\footnote{at least as far as
the dominant channels are concerned.} in the replacement of the photon by a gluon. 
This replacement is indeed not expected to change results concerning the polarisations
of the particles produced.

\section{Discussion of the results at NNLO$^\star$}

First, let us stress that although the uncertainty associated with the choice of the cut 
$s_{ij}^{\rm min}$ is somewhat larger than at NLO$^\star$,  it 
is nevertheless smaller than the one attached to the mass, the renormalisation scale and the factorisation 
scale~\footnote{Note that the uncertainty associated the choice of $\mu_f$ has a 
negligible impact on the final results compared to the other theoretical uncertainties.}. The latter dependence is
expected: on the one hand, we miss the virtual part at low $P_T$ where it is sizable and where it is expected to reduce
the renormalisation-scale dependence;
on the other hand, the contributions  dominating at large $P_T$ are directly sensitive to the fourth power of $\alpha_S$.
Yet, the dependence is smaller than in the inclusive case~\cite{Artoisenet:2008fc} where five powers of $\alpha_S$ are involved.

Second, we find that the subprocess 
$gg \to \Q+\gamma+gg$ dominates, providing with more than two thirds of the whole
yield in the $J/\psi$ case. In addition, we have checked that this fraction 
is slightly increasing with $P_T$ and only weakly dependent on the value of 
the invariant mass cut-off of light partons $s_{ij}^{\rm min}$, removing the 
collinear and infrared divergences of $gg \to \Q+\gamma+gg$. Another important 
observation is that the size of the yield from $gg \to \Q+\gamma+gg$  does
 not vary much when $s_{ij}^{\rm min}$ is changed from 2.25 to 9 GeV$^2$ in the $J/\psi$ case
for instance. This indicates that those divergences are not
--after being cut--  artificially responsible for a large part of the NNLO$^\star$ yield.

Third, it is likely that the  largest part of this contribution is not from
gluon fragmentation topologies, but rather from double $t$-channel gluon exchange ones\footnote{In the inclusive case, those 
have been previously discussed~\cite{kt} in the $k_t$ factorisation formalism. See~\cite{Baranov:2008yk} for a recent
application to $\Upsilon$ hadroproduction.}, keeping in
mind that such a decomposition in terms of the corresponding Feynman graphs is not gauge invariant. 
There are a couple of indications supporting this:
\begin{itemize}
\item Such fragmentation contributions would be expected to provide transverse quarkonia if there were neither corrections
due to the off-shellness of the fragmenting gluons  nor possible spin-flip contributions when radiated gluon energies 
(in $\Q$ rest frame)
are not small compared to the heavy-quark mass. While a depolarisation (in the helicity frame) is possible, it would be still far
from the longitudinal polarisation computed here for $\Q$ produced via $gg \to \Q+\gamma+gg$. 
\item Processes such as $q q' \to \Q +\gamma+q q'$, which proceed uniquely via  double $t$-channel gluon exchange, have
the same  $P_T$ dependence as the process $gg \to \Q+\gamma+gg$ and the difference in normalisation
is naturally accounted for by colour factors and
the smaller value of the quark PDF compared to the gluon one at low $x$.  
Another similarity with $gg \to \Q+\gamma+gg$ is that 
the polarisation of the  yield from $qq' \to \Q +\gamma+qq' $ is strongly longitudinal, 
for $P_T$ larger than 5 GeV, as observed for the full NNLO$^\star$ yield dominated by $gg \to \Q+\gamma+gg$. 
\end{itemize}

The previous discussion can in fact be extended to the inclusive case studied in ~\cite{Artoisenet:2008fc}. Indeed, 
the results then obtained at NNLO$^\star$ appeared to be significantly higher than the ones using the 
fragmentation approximation. Further, the polarisation of the yield from $gg \to \Q+ggg$ processes happened
to become more and more longitudinal at large $P_T$ in contradiction with the expected polarisation
from a fragmentation channel. This lets us think that, for both processes $\Q+\gamma$ and $\Q+X$ at NNLO$^\star$, the yield
may mainly come from double $t$-gluon channel exchanges appearing for the first time at this order.

A careful kinematical analysis of the yield from $gg \to \Q+\gamma+gg$ would certainly be helpful. 
Yet,  it would be highly computer-time demanding, especially 
to obtain a distribution\footnote{The evolution of the distribution for different $P_T$ would be even better.}
of the relative momentum between the $J/\psi$ and its 
closest gluon precise enough to unequivocally attribute the most part of the yield to the double $t$-channel 
gluon exchange channels rather than to the fragmentation ones. Indeed, it concerns 
the most complicated process with a couple of hundreds of diagrams. This work is left for
a future analysis and would certainly be expediently done along with the one for the inclusive case.

\section{Phenomenology}

\subsection{Photon detectability}

In order to detect the photon,  we evidently have to impose that it possesses a finite transverse momentum 
for it not to go in the beam pipe and that it is isolated to avoid misidentifications with Bremsstrahlung radiations. 
Yet, isolation criteria highly depend on the detector potentialities. The determination of
optimum values for $P^\gamma_{T,min}$ and for an isolation criterion for the $\gamma$ improving
the signal over background ratio is beyond the scope of this theoretical analysis. We have therefore decided
to apply minimal values for those constraints, which would be further increased after a full detector simulation.
Along the same lines, it is worth noting here that isolation cut for CO mediated signal (a priori suppressed, as discussed
below) can only be imposed in such simulation. Indeed, such a cut requires a simulation of the hadronic activity
following the discolouration of the CO pairs as done \eg~in \cite{Kraan:2008hb}.

At LO, the $P_T$ requirement is trivially satisfied for a $\Q$ with a finite $P^\Q_T$ since $P^\gamma_T$ is 
balancing $P^\Q_T$. As discussed by Li and Wang~\cite{Li:2008ym}, this is not automatically the case at NLO 
and a minimum $P^\gamma_T$ cut has to be applied and it affects the yield vs $P^\Q_T$ up to roughly three 
times the value of this cut. Typically the cut $P^\gamma_{T,min}=3$ GeV applied here affects the yield
up to $P^\Q_T=7-10$ GeV. As regards the isolation cut, we excluded any events with partons ($q$,$g$) within
a cone $\Delta R=\sqrt{\Delta \eta^2+\Delta \phi^2}=0.1$ from the 
photon~\cite{Jacobs}\footnote{On the way, note for completeness that this cut avoids
the QED singularities that may appear in the processes $qg\to qg\gamma \Q$
when the $\gamma$ is emitted by the external quark. Those are anyhow sub-dominant topologies
of suppressed quark-gluon initiated contributions and could have safely been  neglected.}. As noted above,
those are minimal cuts.

In our case an analysis of different $P^\gamma_T$-cut values is less relevant than in~\cite{Li:2008ym} for two reasons: first, 
our computation cannot be reliably extended to low $P_T$ 
and, second, the main contributions --the double $t$-channel gluon exchange and gluon-fragmentation channels-- create a $\gamma$ 
with a similar momentum as that of the $\Q$ (see the graphs of \cf{diagrams} (d) \& (e)). Overall, such a kinematical cut
insuring its detectability does not affect the NNLO$^\star$ yield for $P_T$ larger than 10 GeV.

Beside the problem of the photon detectability, we have to make sure that other processes will not contribute to 
the yield of $\Q +\gamma$. In the $J/\psi$ case, we expect the non-prompt background to be properly subtracted (by 
vertex-displacement method for instance). However, in this case, the ratio ``non-prompt over prompt'' at very large $P_T$ 
is expected to be lower than in the inclusive case since the photon emission is required for both processes, whereas 
in the inclusive case the $B$ feed-down can proceed at a lower order in $\alpha_S$. Similarly, the $\chi_Q$ feed-down will not 
be significant (except in the region where the invariant mass of the pair $\Q+\gamma$ is close to the mass of the $\chi_Q$)
since suppressed by $v^2$ and by the branching while having the same $\alpha_s$ suppression than for the $J/\psi$
 for similar topologies. As regards the feed-down of the radially excited states $^3S_1$, they
are readily accounted for by constant multiplicative factors to the cross section of the state fed in : 
$\sim 1.4$ for the $\psi(2S)$ into $J/\psi$,  $\sim 1.1$ for both the $\Upsilon(2S)$ into $\Upsilon(1S)$ and 
$\Upsilon(3S)$ into $\Upsilon(2S)$~\footnote{The other feed-down factors can be neglected in view of the low accuracy at
which we know $|R_\Q(0)|^2$ and the respective branchings from which they are derived.}.

\subsection{Colour-octet yield}

Now, let us discuss the possible colour-octet contributions, ignoring first the modifications, 
induced by the QCD corrections to CO and CS contributions, of the 
CO LDME extracted from the Tevatron data.  
As discussed in~\cite{Li:2008ym}, a quick comparison with the results 
obtained in~\cite{Kniehl:2002wd} shows that the NLO CS and the LO CO yields are of the same order at the LHC, 
with roughly the same $P_T$ dependence. 

Yet, contrary to what is claimed in~\cite{Li:2008ym}, higher-QCD corrections are expected to
impact more on the CO partonic matrix elements in $\Q +\gamma$ than in the inclusive case~\cite{Gong:2008ft}, for which
 NLO corrections 
seem unimportant at large $P_T$. 
Indeed, the difference between the two cases lies in the fact that in $\Q +\gamma$ $P_T^{-4}$ topologies, such as fragmentation ones,
 are not opened at the order $\alpha\alpha_S^2$ as studied in~\cite{Kniehl:2002wd} (only one initiated by quark, 
and thus sub-dominant at low $x$, $q\bar q\to \gamma + (Q\bar Q)^{[8]}$,  with a quark in the $t$-channel shows a $P_T^{-4}$ scaling).

In the light of the NNLO$^\star$ results, we know that when the real $\alpha_S^4\alpha$ contributions 
are taken into account, the CS contributions will be increased by about one order of magnitude compared
to the CS at NLO. To be sure that CS contributions are then dominant over CO ones at higher orders, 
we have to check that fragmentation contributions from CO opening at $\alpha\alpha_S^3$ 
are small. 

This is indeed the case since they are mediated by either  $^1S_0^{[8]}$ or $^3P_0^{[8]}$ $C=+1$ CO
states and their corresponding LDMEs are know to be severely constrained when higher-QCD corrections are taken into account
for the analysis of the inclusive data at low- and mid-$P_T$ at the Tevatron~\cite{Gong:2008ft,Kniehl:1998qy}, at various 
fixed-target experiments~\cite{Maltoni:2006yp} and more recently at the $B$ factories~\cite{Ma:2008gq,Gong:2009kp}.
Finally, a further suppression is expected from the increase of the CS yield from NLO and NNLO$^\star$ 
contributions~\cite{Artoisenet:2008fc}.
Using an upper conservative value of the LDME combination $M_{k}^{J\psi}= \langle 0| {\cal O}^{J/\psi}[^1S_0^{[8]}]|0\rangle
+ \frac{k}{m_c^2}\langle 0| {\cal O}^{J/\psi}[^3P_0^{[8]}]|0\rangle =1 \times 10^{-2}$ GeV$^3$ ($k=3.5$),
we analysed the two extreme cases
($\langle 0| {\cal O}^{J/\psi}[^3P_0^{[8]}]|0\rangle=0$ (case I) and $\langle 0| {\cal O}^{J/\psi}[^1S_0^{[8]}]|0\rangle=0$ (case II))
to evaluate those CO $\alpha\alpha_S^3$ contributions via 
$pp\to j +  (c\bar c)^{8,C=+1}+ \gamma $ using the same parameter values as the central NLO curve and $s^{\rm min}_{ij}=m_c^2$.
At $P_T\simeq 10$ GeV, the case I gives a contribution about 5 times smaller than the central $\alpha\alpha_S^3$ (NLO) CS values 
while the case II is further below\footnote{This indicates on the way that
this process is sensitive to a different linear combination of $C+1$ CO LDME than the inclusive production.} by another factor of 5.
At $P_T\simeq 40$ GeV, case I is less than a factor of 2 larger than the 
central NLO values, thus still clearly below the NNLO$^\star$ yield, while case II is still a bit suppressed and
comparable to the central NLO values. A dedicated analysis including a global fit of the $C=+1$ CO LDMEs
and taking into account DGLAP radiation of the fragmenting gluon is beyond the scope of this work but
would confirm that $\alpha\alpha_S^3$ CO corrections to $\Q +\gamma$ matter only at large $P_T$ and are most likely not large
enough to challenge the dominance of CS contributions to this process.

This dominance is expected since the photon has to be emitted by quarks. 
If the hard scattering part is $gg\to gg$, the photon can only be emitted by the heavy quarks with topologies
similar to CS channels.  At $\alpha_S^2\alpha$, the gluon-fusion CO channels
do not scale like $P_T^{-4}$. 
One has to go to  $\alpha_S^3\alpha$ to have  the first  gluon 
fragmentation channels initiated by gluon fusion. In this case, as discussed above,
the gluon fragments into a photon and a CO $C=+1$ (whose LDMEs are severely constrained)
\footnote{
If one required that $P^\gamma_T$ exactly balance $P^\Q_T$ in order to select the LO CSM, aiming at the extraction 
of the gluon distributions for instance, one would have to refine the analysis to take into account NNLO$^\star$ 
corrections in the CS channels and the CO yield~\cite{Mathews:1999ye} which would then be dominated by 
$gg\to g^\star\to (Q\bar Q)^{[8]}+\gamma$ via $^1S_0^{[8]}$ or $^3P_J^{[8]}$ if the $C=+1$ CO channels are
not too much suppressed.}. Last but not least, contrary to the double $t$-channel gluon exchanges dominating the CS yield, 
the latter fragmentation CO topologies will systematically produce nearly collinear $\gamma$ and 
hadrons from the discolouration of the CO state. Those events are therefore likely to
be rejected by experimental photon-isolation cuts.

Finally, while contributions from $s$-channel cut to $J/\psi$ production could appear in the low $P_T$ region as in the 
inclusive case~\cite{Haberzettl:2007kj} (the final state gluon of the $gg\to J/\psi g$ being simply replaced by a photon), 
in the large $P_T$ region, the colour-transfer-enhancement mechanism discussed in~\cite{Nayak:2007mb} is not 
expected to matter for the present process. In the inclusive case,  $gg\to J/\psi+c \bar c$ may account for a significant part
of the yield at very large $P_T$. This indicates that topologies for which colour transfer could occur are not much suppressed
and those could impact on the inclusive yield.
In the present case, we have checked that the contribution 
to $J/\psi+\gamma$ from the partonic process $gg\to J/\psi+ c\bar c+\gamma$ is sub-dominant and found that 
it is one order of magnitude smaller
than the process from $gg\to J/\psi+\gamma+gg$ with the same $P_T$ dependence between
20 and 50 GeV. It is therefore quite unlikely that colour transfers, acting on topologies
with at least 3 heavy quarks, be visible in $J/\psi+\gamma$ yields.

\section{Conclusion}

In conclusion, we have computed the real next-to-next-to-leading order QCD
contributions to the hadroproduction of a $J/\psi+\gamma$ and $ \Upsilon+\gamma$ via colour singlet 
transitions along the same lines as~\cite{Artoisenet:2008fc} for the inclusive case
and argued that it provided a first reliable evaluation of the corresponding yield at NNLO accuracy.

Indeed, prior to that,  we have shown that the full NLO evaluation of Li and Wang~\cite{Li:2008ym}
is very accurately reproduced for $P_T>5$ GeV by the sole evaluation of the real emission 
contributions up to order $\alpha^3_S\alpha$, namely the NLO$^\star$. 
In the inclusive case~\cite{Artoisenet:2008fc}, a similar observation was made and 
motivated the study of the NNLO$^\star$ contributions to evaluate the yield at NNLO accuracy.
We have thus reach a completely independent, but similar, conclusion for the production
of a quarkonium in association with a photon.

A priori, integrating the amplitudes for such
real emission contributions leads to divergences in some phase-space
regions. Those can be avoided by imposing a minimal invariant mass between
the external light partons, while in a complete calculation those divergences
 are canceled by the virtual corrections.
Yet, at NLO, the latter scale  as $P_T^{-8}$ and, at NNLO,
 as $P_T^{-6}$, hence are suppressed compared to  respectively
the real NLO part scaling as $P_T^{-6}$ and the real
 NNLO part scaling as $P_T^{-4}$. This explains why 
we can neglect the virtual corrections at mid and large $P_T$ and use a cut-off 
on which the results become insensitive when $P_T$ grows,
and also why the sole NLO$^\star$ reproduces the full NLO accurately.

We have also shown that the differential cross section in $P_T$ at 
NNLO$^\star$ is one order of magnitude larger than 
at NLO for large $P_T$  --as expected from their $P_T$ scaling--. We 
have also identified
the process responsible for the most part of it, i.e. $gg \to \Q +\gamma +gg$.
Although, the distinction between the gluon-fragmentation and
the double $t$-channel gluon exchange graphs cannot be carried out in gauge invariant way, we
provided some hints that the second type of topology is dominating, similar 
to the inclusive case~\cite{Artoisenet:2008fc}. One of this hint is the polarisation
of the quarkonia produced in association with a photon.

Indeed, we have computed the polarisation parameter
$\lambda$ of the NNLO$^\star$ yield which is negative,
indicating a longitudinally polarised yield. This confirms 
the trend observed at NLO and hints at the dominance 
of double $t$-channel gluon exchange contributions.

When the NNLO$^\star$ contributions are incorporated in the CS yield, it becomes one order of magnitude
larger than the potential CO yield, which would mainly produce transversally polarised $\Q$. 
The measurements of the cross section for the production of $\Q+\gamma$ 
would directly measure  the size of the CS without being
sensitive to the non-perturbative CO parameters. This is a complementary case to the study
of $J/\psi + c \bar c$ and $\Upsilon + b \bar b$ as discussed in~\cite{Artoisenet:2007xi,Artoisenet:2008tc,Lansberg:2008gk}.

Similarly to the analysis of the inclusive case~\cite{Artoisenet:2008fc},
this analysis cannot be extended to too low $P_T$, where the
approximations on which it is based no longer hold. One way to 
improve the predictions could be achieved by merging the matrix elements with
parton showers using one of the approaches available in the
literature~\cite{Alwall:2007fs}, or by performing an
analytic resummation~\cite{Berger:2004cc}.

Finally, the results presented here strongly support the procedure used for the very first evaluation of
the inclusive yield at NNLO accuracy~\cite{Artoisenet:2008fc} and which showed an agreement with the Tevatron measurements.
This in turn confirms that  much better Monte Carlo simulations than the ones based
on matrix elements for CS channels at LO only are now possible  at mid and large $P_T$. 
This could be achieved using NLO$^\star$ and NNLO$^\star$ partonic matrix elements generated 
by MadOnia~\cite{Artoisenet:2007qm,Alwall:2008pm}, for the process studied here
$pp \to \Q +\gamma +X$, but also for the inclusive measurements to be performed at the LHC.

\section*{Acknowledgements}
We are thankful to  R.~Li and J.~X.~Wang for providing their data points and for explanations on their work. 
We are grateful to  P. Artoisenet, S.J. Brodsky, J.R. Cudell and P.~Hoyer for comments on the manuscript, 
to O.~Mattelaer for his very helpful
suggestions and to L.~Dixon, B. Hippolyte, P. Jacobs, F. Maltoni and V.N. Tram for useful discussions or comments.
This work is supported in part by a Francqui fellowship of the Belgian American Educational Foundation and
by the U.S. Department of Energy under contract number DE-AC02-76SF00515.

\end{document}